\documentclass[aps,prl,twocolumn,amsmath,amssymb,reprint,groupedaddress,preprintnumbers]{revtex4}

\usepackage{enumitem}
\usepackage{graphicx}
\usepackage{hyperref}

\newcommand{\HBS}{\mathbb S}
\newcommand{\M}{j}

\newcommand{\zt}{\zeta_3}

\newcommand{\zf}{\zeta_5}

\newcommand{\cN}{{\cal N}}

\begin{document}


\preprint{DESY~20--033\hspace{13.5cm}ISSN 0418--9833\phantom{XXXXXX}}
\preprint{October 2020\hspace{17.3cm}}

\title{Nonplanar Cusp and Transcendental Anomalous Dimension at Four Loops\\
in $\cN=4$ Supersymmetric Yang-Mills Theory}


\author{B.~A.~Kniehl}
\email[]{kniehl@desy.de}
\affiliation{{II.} Institut f{\"u}r Theoretische Physik,
Universit{\"a}t Hamburg,
Luruper Chaussee 149, 22761 Hamburg, Germany}

\author{V.~N.~Velizhanin}
\email[]{velizh@thd.pnpi.spb.ru}
\affiliation{Theoretical Physics Division, NRC ``Kurchatov Institute,''
Petersburg Nuclear Physics Institute,
Orlova Roscha, Gatchina, 188300 St.~Petersburg, Russia}

\date{\today}

\begin{abstract}
  We compute the nonplanar contribution to the universal anomalous dimension of
  the SU(4)-singlet twist-two operators in $\cN=4$ supersymmetric Yang-Mills theory at four loops through Lorentz spin 18. From this, we numerically
  evaluate the nonplanar contribution to the four-loop lightlike cusp anomalous dimension and derive the transcendental $\zeta_3$ and $\zeta_5$ parts of the universal anomalous dimension for arbitrary Lorentz spin in analytic form.
  As for the lightlike cusp anomalous dimension and the $\zeta_5$ part of the
  universal anomalous dimension, we confirm previous results.  
\end{abstract}


\maketitle


The AdS/CFT correspondence
\cite{Maldacena:1997re,Gubser:1998bc,Witten:1998qj},
also known as holographic duality, has been one of the most active and
tantalizing research topics in high-energy theory over the past two decades. 
This implies that quantum gravity in anti--de Sitter space, with constant
negative curvature, is equivalent to a lower-dimensional nongravitational
quantum field theory of conformal type, $\cN=4$ supersymmetric Yang-Mills (SYM) theory,
living on the boundary of that gravitational space.
The AdS/CFT correspondence has led to a plethora of intriguing physical
insights and powerful novel methods of calculation~\cite{Witten:2003nn,Cachazo:2004kj,Britto:2004nc,Britto:2005fq,Britto:2004ap,Beisert:2005fw,Bern:2005iz,Beisert:2006ez,Alday:2007hr,Bern:2008qj,ArkaniHamed:2010kv,Beisert:2010jr,Gromov:2013pga}.
The latter allow us to solve longstanding problems not only in supersymmetric
toy models, but also in real theories of nature, such as quantum chromodynamics
(QCD)~\cite{Berger:2009zg,Berger:2010zx,Bern:2013gka}.

So far, investigations of the AdS/CFT correspondence have largely been confined
to the planar limit, in which Feynman diagrams of planar topologies contribute,
while nonplanar topologies are far more difficult to tackle. 
It is obviously of paramount interest to go beyond the planar limit, as this
will allow us to significantly deepen and consolidate our understanding of the
AdS/CFT correspondence and to access as-yet unexplored regions of~it.

Quantities of key interest include the anomalous dimensions of the operators,
composed of the quantum fields of $\cN=4$ SYM theory, that are of leading
twist, twist two, and are singlets under the internal symmetry group SU(4).
These operators are sorted by their Lorentz spin $j$, which counts
the covariant derivatives, and are multiplicatively renormalized, sharing the
same universal anomalous dimension $\gamma_{\mathrm {uni}}(\M)$, which just
depends on $\M$.
Nonplanar contributions to the latter can be obtained by directly computing,
by means of advanced computerized methods, the relevant Feynman diagrams in
perturbation theory in powers of the gauge coupling~$g$.

The study of the renormalization of composite operators in $\mathcal{N}=4$ SYM
theory has led to the discovery of the relation of this problem with exactly
solvable models~\cite{Minahan:2002ve}.
The integrability in the planar limit was intensively studied and established from both sides of the AdS/CFT correspondence (see Ref.~\cite{Beisert:2010jr} for a review and  Refs.~\cite{Gromov:2013pga,Gromov:2014caa} for the recently developed quantum spectral curve approach).
In the nonplanar case, integrability-based methods have been considered, in general, in Refs.~\cite{Bargheer:2017nne,Bargheer:2018jvq}.
Nonplanar contributions to anomalous dimensions serve as a welcome laboratory
for stringent tests of the ideas and models thus proposed.
This provides a strong motivation for our present work. 

Once a general result for the universal anomalous dimension is established, it
is interesting to study its analytical properties and particular limits.
The most interesting one, $j\to\infty$, yields the lightlike cusp anomalous
dimension $\gamma_{\mathrm{cusp}}$ \cite{Korchemsky:1988si}, which can be computed by alternative methods, too.
The planar part of $\gamma_{\mathrm{cusp}}$ has been found to all orders a long
time ago, through the asymptotic Bethe {\it Ansatz} equation \cite{Beisert:2006ez}. 
Recently, its nonplanar part has been established through four loops, at
$\mathcal{O}(g^8)$, via the Sudakov form factor, numerically in
Refs.~\cite{Boels:2015yna,Boels:2017skl} and analytically in
Ref.~\cite{Huber:2019fxe}, and via the lightlike polygonal Wilson loops, again
analytically, in Ref.~\cite{Henn:2019swt}.
At four loops in QCD, at $\mathcal{O}(\alpha_s^4)$ in the strong-coupling
constant $\alpha_s$, the quark cusp anomalous dimension in the planar limit has
been found via the quark form factor in Ref.~\cite{Lee:2016ixa}, its
contribution with quartic fundamental color factor has been obtained, again via
the quark form factor, in Ref.~\cite{Lee:2019zop}, and the complete quark and
gluon cusp anomalous dimensions have been established via their counterpart in
${\mathcal N}=4$ SYM theory in Ref.~\cite{Henn:2019swt} and via the massless
quark and gluon form factors in Ref.~\cite{vonManteuffel:2020vjv}.

Explicit knowledge of $\gamma_{\mathrm {uni}}(\M)$ for general value of $j$ would
unfold the nonplanar anatomy of the anomalous dimensions in $\cN=4$ SYM
theory.
A possible avenue to this goal is to evaluate $\gamma_{\mathrm {uni}}(\M)$ for
as many values of $j$ as possible and to try and extract from this the general
result.
In $\cN=4$ SYM theory, nonplanarity appears for the first time at
$\mathcal{O}(g^8)$.
In Refs.~\cite{Velizhanin:2009gv,Velizhanin:2010ey,Velizhanin:2014zla}, the
nonplanar contributions to $\gamma_{\mathrm {uni}}(\M)$ at $\mathcal{O}(g^8)$
were analytically calculated for the first three nontrivial values $j=4,6,8$.
Recently, these results have been confirmed and extended to $j=10$ applying the
method of asymptotic expansions to the four-point functions of length-two
half--Bogomol'nyi-Prasad-Sommerfield operators~\cite{Fleury:2019ydf}.
The purpose of this Letter is to push this endeavor as far as possible, which
turns out to be through $j=18$, thanks to cutting-edge technology and computing
power. 
We will thus be able to reconstruct the general coefficients of $\zeta_3$ and
$\zeta_5$ and to obtain an independent numerical result for
$\gamma_{\mathrm{cusp}}$ at $\mathcal{O}(g^8)$.
The former is new, and the latter confirm previous findings in
Ref.~\cite{Velizhanin:2009gv} and
Refs.~\cite{Boels:2015yna,Boels:2017skl,Henn:2019swt,Huber:2019fxe},
respectively.

Specifically, the set of local, gauge-invariant, SU(4)-singlet, twist-two
operators of definite Lorentz spin $j$ in $\cN=4$ SYM theory reads:
\begin{eqnarray}
\mathcal{O}_{\mu _{1},...,\mu _{\M}}^{\lambda } &=&\hat{S}
\bar{\lambda}_{i}^{a}\gamma _{\mu _{1}}
{\mathcal D}_{\mu _{2}}\cdots{\mathcal D}_{\mu _{\M}}\lambda ^{a,i}\,, \label{qqs}\\
\mathcal{O}_{\mu _{1},...,\mu _{\M}}^{g} &=&\hat{S} G_{\rho \mu_{1}}^{a}{\mathcal
  D}_{\mu _{2}} {\mathcal D}_{\mu _{3}}\cdots{\mathcal D}_{\mu _{\M-1}}
G_{\mu_{\M}}^{a,\rho}\,,
\label{ggs}\\
\mathcal{O}_{\mu _{1},...,\mu _{\M}}^{\phi } &=&\hat{S}
\bar{\phi}_{r}^{a}{\mathcal D}_{\mu _{1}} {\mathcal D}_{\mu _{2}}\cdots{\mathcal
D}_{\mu _{\M}}\phi^{a,r}\,,\label{phphs}
\end{eqnarray}
where the spinors $\lambda_{i}$ refer to the gauginos, the field strength
tensor $G_{\mu\nu}$ to the gauge fields, $\phi_{r}$ are the complex scalar
fields of extended supersymmetry, and ${\mathcal D}_\mu$ are covariant
derivatives.
The indices $i=1,\ldots,4$ and $r=1,\ldots,3$ refer to the SU(4) and
SO(6)${}\simeq{}$SU(4) groups of internal symmetry, respectively.
The symbol $\hat{S}$ implies a symmetrization of the respective tensor in the Lorentz
indices $\mu_{1},...,\mu _{\M}$ and a subtraction of all its possible traces. 
As mentioned above, these operators form the multiplicatively renormalized operators, whose
anomalous dimensions are expressed through the so-called universal
anomalous dimension up to integer argument shifts~\cite{Kotikov:2002ab},
\begin{equation}
  \gamma_{\mathrm {uni}}(\M)=\sum_{n=1}^{\infty}\gamma_{\mathrm {uni}}^{(n-1)}(\M)\,g^{2n},
  \qquad g^2=\frac{\lambda}{16\pi^2}\,,
\end{equation}
where $\lambda=g^2_{\mathrm {YM}}N_c$ is the 't~Hooft coupling constant.

In the planar limit, $\gamma_{\mathrm {uni}}(\M)$ is analytically known for
arbitrary $j$ through seven loops \cite{Kotikov:2003fb,Kotikov:2004er,Bajnok:2008qj,Lukowski:2009ce,Marboe:2014sya,Marboe:2016igj} and
for special values of $j$ even through ten loops \cite{Fiamberti:2007rj,Bajnok:2008bm,Velizhanin:2008jd,Bajnok:2009vm,Bajnok:2012bz,Leurent:2012ab,Leurent:2013mr,Marboe:2014gma}, e.g.\ for $\M=4$ corresponding to the Konishi operator.
In the latter case, we quote the result through four loops
\cite{Fiamberti:2007rj,Bajnok:2008bm,Velizhanin:2008jd} here
\begin{eqnarray}
  \gamma_{\mathrm{Konishi,planar}}&=&
  \gamma_{\mathrm{uni,planar}}(4)
    =12g^2-48g^4+336g^6\nonumber\\
  &&\hspace*{-10mm}{}+96g^8(-26 + 6 \zeta_3 - 15 \zeta_5)+{\mathcal O}(g^{10})\,.
 \end{eqnarray}

As for the nonplanar contributions to $\gamma_{\mathrm {uni}}(\M)$ at
$\mathcal{O}(g^8)$, the state of the art is given by
\cite{Velizhanin:2009gv,Velizhanin:2010ey,Velizhanin:2014zla}
\begin{eqnarray}
\gamma_{\mathrm {uni,np}}^{(3)}(4)&=&
- 360 \zf \frac{48}{N_c^2}\,,
\label{guniM4}\\[2mm]
\gamma_{\mathrm {uni,np}}^{(3)}(6)&=&
\frac{25}{9}(21 + 70  \zt - 250 \zf)  \frac{48}{N_c^2}\,,
\label{guniM6}\\
\gamma_{\mathrm {uni,np}}^{(3)}(8)& =&
\frac{49}{600}(1357 + 4340 \zt - 11760 \zf) \frac{48}{N_c^2}\,,
\label{guniM8}
\end{eqnarray}
where we have pulled out common factors.
If such a factorization were preserved for the higher $j$ values, this could
considerably simplify the procedure of finding the general form of
$\gamma_{\mathrm {uni,np}}^{(3)}(j)$.
In fact, the prefactors in Eqs.~\eqref{guniM4}--\eqref{guniM8} resemble the
harmonic sums $\sum_{i=1}^{j-2}\frac{1}{i}$ for $j=4,6,8$, with values $3/2$,
$25/12$, $49/20$, and harmonic sums are also expected to appear as building
blocks of $\gamma_{\mathrm {uni,np}}^{(3)}(j)$, as explained below.

\begin{figure}
\begin{center}
\includegraphics[width=0.47\textwidth]{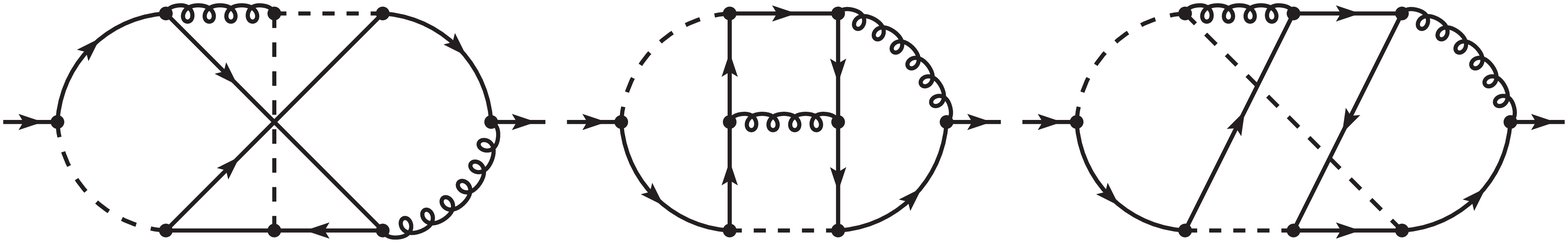}
\caption{Typical Feynman diagrams contributing to $\gamma_{\mathrm {uni},\mathrm{np}}^{(3)}(\M)$.
The operators are inserted in the lines or gauge vertices.}
\label{fig:diagrams} 
\end{center}
\end{figure}

In this Letter, we extend Eqs.~\eqref{guniM4}--\eqref{guniM8} by the next five
terms.
Our computational procedure is similar to
Refs.~\cite{Velizhanin:2009gv,Velizhanin:2010ey,Velizhanin:2014zla}.
We work in the programming language {\sc form} \cite{Vermaseren:2000nd}.
Specifically, we generate all the contributing Feynman diagrams with
{\sc diana} \cite{Tentyukov:1999is} based on {\sc qgraf}
\cite{Nogueira:1991ex},
evaluate the color traces with {\sc color} \cite{vanRitbergen:1998pn},
reduce the occurring scalar integrals to the master integrals of
Ref.~\cite{Czakon:2004bu} with the custom-made program package {\sc bamba}
based on the Laporta algorithm~\cite{Laporta:2001dd}, and reduce the
propagator-type diagrams to fully massive tadpole diagrams using
infrared rearrangement \cite{Vladimirov:1979zm} (see also
Refs.~\cite{Misiak:1994zw,Chetyrkin:1997fm,Czakon:2004bu} for details).
Typical Feynman diagrams are depicted in Fig.~\ref{fig:diagrams}.
This setup allows us to proceed to $j=10$ only.
Further progress is enabled by the recently developed package {\sc forcer}
\cite{Ruijl:2017cxj} based on {\sc form} \cite{Vermaseren:2000nd}, which
was also used to compute anomalous dimensions of twist-two operators in QCD
\cite{Moch:2017uml,Moch:2018wjh}.
We may thus proceed to $j=18$, where about one year of running on the modern
high-performance computing facilities available to us is required.  
Altogether, we have
\allowdisplaybreaks[1]
\begin{align}
\gamma_{\mathrm {uni,np}}^{(3)}(10) & =
\bigg(\frac{220854227}{1411200}+\frac{27357}{56}\zt
-\frac{579121}{490}\zf\bigg)\frac{48}{N_c^2}\,,
\label{guniM10}\\[2mm]
  \gamma_{\mathrm {uni,np}}^{(3)}(12)& =
\bigg(\frac{28337309747461}{144027072000}
+\frac{345385183}{571536}\zt\nonumber\\&\hspace*{15mm}
-\frac{54479161}{39690}\zf\bigg)\frac{48}{N_c^2}
\,,\label{guniM12}\\[2mm]
\gamma_{\mathrm {uni,np}}^{(3)}(14)& =
\bigg(\frac{9657407179406311}{41493513600000}
+\frac{158654990663}{224532000}\zt\nonumber\\&\hspace*{15mm}
-\frac{7399612441}{4802490}\zf\bigg)\frac{48}{N_c^2}
\,,
\label{guniM14}\\[2mm]
\gamma_{\mathrm {uni,np}}^{(3)}(16)& =
\bigg(\frac{74429504651244877}{280496151936000}
+\frac{205108095887}{256864608}\zt\nonumber\\&\hspace*{15mm}
-\frac{1372958223289}{811620810}\zf\bigg)\frac{48}{N_c^2}\,,
\label{guniM16}\\[2mm]
\gamma_{\mathrm {uni,np}}^{(3)}(18)& =
\bigg(\frac{8122582838282649980649377}{27516111512617728000000}\label{guniM18}\\&\hspace*{-10mm}
+\frac{72169501556777041}{81811377648000}\zt
-\frac{5936819760481}{3246483240}\zf\bigg)\frac{48}{N_c^2}\,.\nonumber
\end{align}
Our result for $j=10$ agrees with Ref.~\cite{Fleury:2019ydf}.
In contrast to Eqs.~\eqref{guniM4}--\eqref{guniM8}, it is not possible to
extract common factors in Eqs.~\eqref{guniM10}--\eqref{guniM18}, so that
our expectations regarding factorization have to be dropped.

Equipped with the information contained in Eqs.~\eqref{guniM4}--\eqref{guniM18},
we now try to reconstruct the general form of $\gamma_{\mathrm {uni,np}}^{(3)}(j)$,
i.e.\ to determine the $j$ dependence of the coefficients of $\zeta_5$ and
$\zeta_3$ and the rational reminder in the {\it Ansatz} 
\begin{eqnarray}\label{HSZ5ResALL}
  \gamma_{{\mathrm{uni}},{\mathrm{np}}}^{(3)}(\M)&=&\Big[\gamma_{{\mathrm{uni}},{\mathrm{np}},\zf}^{(3)}(\M)\,\zf+\gamma_{{\mathrm{uni}},{\mathrm{np}},\zt}^{(3)}(\M)\,\zt
  \nonumber\\&&{}
+\gamma_{{\mathrm{uni}},{\mathrm{np}},{\mathrm{rational}}}^{(3)}(\M)\Big]\frac{48}{N_c^2}\,.
\end{eqnarray}
For this purpose, we adopt a powerful method based on number theory which has
been proposed in Ref.~\cite{Velizhanin:2010cm} and successfully applied to
the reconstruction of anomalous dimensions in $\mathcal{N}=4$ SYM theory
\cite{Velizhanin:2013vla,Marboe:2014sya,Marboe:2016igj} and QCD in Ref.~\cite{Velizhanin:2012nm} and after in Refs.~\cite{Moch:2014sna,Moch:2017uml}.
This method is based on the assumption that
$\gamma_{{\mathrm{uni}},{\mathrm{np}},\zf}^{(3)}(\M)$,
$\gamma_{{\mathrm{uni}},{\mathrm{np}},\zt}^{(3)}(\M)$, and
$\gamma_{{\mathrm{uni}},{\mathrm{np}},{\mathrm{rational}}}^{(3)}(\M)$
in Eq.~\eqref{HSZ5ResALL} are linear combinations of certain basis functions
with certain coefficients.
As for the basis functions and coefficients, we are guided by several
heuristic observations. 

As for the basis functions, in the case of anomalous dimensions of twist-two
operators in $\cN=4$ SYM theory, these are known to be generalized harmonic
sums, defined as \cite{Vermaseren:1998uu,Blumlein:1998if}
\begin{eqnarray} 
S_{a_1,\ldots,a_n}(M)&=&\sum^{M}_{j=1} \frac{[\mbox{sign}(a_1)]^{j}}{j^{\vert a_1\vert}}
\,S_{a_2,\ldots,a_n}(j)\,,\label{vhs}
\end{eqnarray}
where the indices $a_1,\ldots,a_n$ may take all (positive and negative) integer
values, except for $-1$.
The weight or \textit{transcendentality} $\ell$ of the sum $S_{a_1,\ldots,a_n}$ is
defined as the sum of the absolute values of its indices,
$\ell=\vert a_1 \vert +\cdots +\vert a_n \vert$, 
and the weight of a product of generalized harmonic sums is equal to the sum of
their weights.

For twist-two operators, there is an additional simplification, thanks to the
so-called generalized Gribov-Lipatov reciprocity
\cite{Gribov:1972ri,Gribov:1972rt,Dokshitzer:2005bf,Dokshitzer:2006nm}, which
reflects the symmetry of the underlying processes under the crossing of
scattering channels.
As a consequence, the harmonic sums can enter the anomalous dimensions only in
the form of special combinations satisfying the above-mentioned property by
themselves.
In practice, this allows us to impose restrictions on the choice of basis
functions leaving us with a smaller number of so-called binomial harmonic
sums, defined as \cite{Vermaseren:1998uu}
\begin{equation}
{\mathbb S}_{a_1,\ldots,a_n}(N)=\sum_{j=1}^{N}(-1)^{j+N}\binom{N}{j}\binom{N+j}{j}S_{a_1,...,a_n}(j)\,.
\label{BinomialSums}
\end{equation}
They only have positive-integer indices, while their transcendentality is the
same as for usual harmonic sums.
There are $2^{\ell-1}$ binomial harmonic sums at transcendentality $\ell$.
While reciprocity has not yet been rigorously proven, counterexamples have not
been encountered either.
In particular, the anomalous dimensions of the twist-two operators in the
planar limit of $\mathcal{N}=4$ SYM theory, which are known through seven
loops, may all be represented in terms of binomial harmonic sums and their
derivatives.
In the planar limit of QCD, the nonsinglet quark anomalous dimensions at four
loops can also be written in terms of binomial harmonic sums and their
derivatives \cite{Moch:2017uml}.
Furthermore, the Feynman diagrams contributing to
$\gamma_{\mathrm {uni},\mathrm{np}}^{(3)}(\M)$ at subleading order in $N_c$ also
contribute to its planar counterpart, as may be seen from
Fig.~\ref{fig:diagrams}. 
These observations suggest that reciprocity should work for the case at hand,
which is also confirmed by a nontrivial self-consistency test within our
calculation, as explained below. 

According to the maximal-transcendentality principle~\cite{Kotikov:2002ab}, the anomalous dimensions of twist-two operators at $\ell$th order in $\mathcal{N}=4$ SYM theory are of transcendentality $2\ell-1$, which is seven for our case of
$\ell=4$.
Thus, $\gamma_{{\mathrm{uni}},{\mathrm{np}},\zf}^{(3)}(\M)$,
$\gamma_{{\mathrm{uni}},{\mathrm{np}},\zt}^{(3)}(\M)$, and
$\gamma_{{\mathrm{uni}},{\mathrm{np}},{\mathrm{rational}}}^{(3)}(\M)$
in Eq.~\eqref{HSZ5ResALL} are of transcendentalities 2, 4, and 7; i.e., they
are composed of 2, 8, and 64 binomial harmonic sums of the respective
transcendentality.

As for the coefficients in front of the basis functions, inspection of the expressions of the $j$-dependent anomalous dimensions that are already known reveals that they are usually small integer numbers.
So, in general, we obtain a system of Diophantine equations. If the number of equations is equal to the number of variables, then we can solve such a system exactly. However, this requires the knowledge of the anomalous dimensions for a large number of fixed $j$ values. Fortunately, the system of Diophantine equations can be solved with the help of special methods from number theory even if the number of equations is less than the number of variables. In fact, we may then apply the Lenstra-Lenstra-Lovasz algorithm~\cite{Lenstra82factoringpolynomials}, which allows us to reduce the matrix obtained from the system of Diophantine equations to a form in which the rows are the solutions of the system with the minimal Euclidean norm.

Equation~\eqref{guniM4} is sufficient to fix the two coefficients in the {\it Ansatz} for
$\gamma_{{\mathrm{uni}},{\mathrm{np}},\zf}^{(3)}(\M)$.
The result,
\begin{equation}\label{HSZ5ResS1}
\gamma_{{\mathrm{uni}},{\mathrm{np}},\zf}^{(3)}(\M) = -40\,{\mathbb S}_{1}^2(\M-2)\,,
\end{equation}
thus obtained a long time ago \cite{Velizhanin:2009gv} has been confirmed
by all subsequent results in Eqs.~\eqref{guniM6}--\eqref{guniM18}.
To determine the eight coefficients in the {\it Ansatz} for
$\gamma_{{\mathrm{uni}},{\mathrm{np}},\zt}^{(3)}(\M)$, we need five input relations.
Using Eqs.~\eqref{guniM4}--\eqref{guniM12}, we find
\begin{eqnarray}\label{HSZ3ResS1}
\gamma_{{\mathrm{uni}},{\mathrm{np}},\zt}^{(3)}(\M)& =& 
8(8\, {\HBS}_4-9\, {\HBS}_{1,3}-3\, {\HBS}_{2,2}-4\, {\HBS}_{3,1}
\nonumber\\&&{}
+4\, {\HBS}_{1,1,2}
+5\, {\HBS}_{1,2,1}-{\HBS}_{2,1,1})\,,
\end{eqnarray}
where $\HBS_{\pmb{a}}=\HBS_{\pmb{a}}(\M-2)$, which is in agreement with
Eqs.~\eqref{guniM14}--\eqref{guniM18}.
By the way, this nicely supports our suggestion that reciprocity should work
for $\gamma_{\mathrm {uni},\mathrm{np}}^{(3)}(\M)$ as well.
At any rate, the eight inputs from Eqs.~\eqref{guniM4}--\eqref{guniM18}
uniquely fix Eqs.~\eqref{HSZ5ResS1} and \eqref{HSZ3ResS1}.
Unfortunately, these inputs do not yet suffice to determine the coefficients
of the 64 binomial harmonic sums of transcendentality 7 in
$\gamma_{{\mathrm{uni}},{\mathrm{np}},{\mathrm{rational}}}^{(3)}(\M)$ beyond all doubt
via the number theoretical procedure outlined above.

Nevertheless, we may exploit the information encoded in
Eqs.~\eqref{guniM4}--\eqref{guniM18} to numerically recover 
the nonplanar contribution to the cusp anomalous dimension with useful
precision.
To this end, we proceed along the lines of Refs.~\cite{Moch:2017uml,Moch:2018wjh} and approximately reconstruct the four-loop splitting function.
We recall that the $n$-loop splitting function $P^{(n)}(x)$ is related to the
anomalous dimension of the respective twist-two spin-$j$ operator, with
$j=2,4,6,\ldots$, by a Mellin transformation,
\begin{equation}
\gamma^{(n)}(j) = - \int_0^1 dx\, x^{j-1} P^{(n)}(x) \,,\label{SP}
\end{equation}
where the negative sign is a standard convention.

In QCD, the diagonal splitting functions at $n$ loops, in general, assume the following form in the limit $x\to1$ \cite{Dokshitzer:2005bf}:
\begin{eqnarray}
  P^{(n-1)}_{kk}(x) &=&
        \frac{A_k^{(n)}}{(1-x)_+}
          + B_k^{(n)}\, \delta (1-x)
          + C_k^{(n)}\,\ln (1-x)
          \nonumber\\
          &&{}+ D_k^{(n)} 
  + \mathcal{O}\big[ (1-x) \ln^{\ell} (1-x) \big]\,,\label{xto1}
\end{eqnarray}
where $k=q,g$ and the $+$ distribution is defined as
usual,
$\int_0^1 dx\,f(x)/(1-x)_+=\int_0^1 dx[f(x)-f(1)]/(1-x)$.
$A_q^{(n)}$ and $A_g^{(n)}$ are the $n$-loop quark and gluon
cusp anomalous dimensions, respectively \cite{Korchemsky:1988si}.
In $\mathcal{N}=4$ SYM theory, the splitting functions, being related to the anomalous dimensions through the Mellin transformation in Eq.~\eqref{SP}, satisfy the maximal-transcendentality principle~\cite{Kotikov:2002ab}, and we may use Eq.~\eqref{xto1}, with $kk$ replaced by $\mathrm{np}$.
Since $C_k^{(n)}$ and $D_k^{(n)}$ can be
predicted from lower-order information \cite{Dokshitzer:2005bf} and
nonplanarity appears for the first time at $n=4$, we have
$C_{\mathrm{np}}^{(4)}=D_{\mathrm{np}}^{(4)}=0$.
Following Refs.~\cite{Moch:2017uml,Moch:2018wjh}, we make {\it Ans\"atze} for
approximations of the splitting function $P^{(3)}_{\mathrm {uni,np}}(x)$, which
consist of
(i) the two large-$x$ parameters $A_{\mathrm{np}}^{(4)}$ and $B_{\mathrm{np}}^{(4)}$,
(ii) two out of the three large-$x$ logarithms $(1-x)\ln^k(1-x)$ with $k=1,2,3$,
(iii) two out of the three small-$x$ logarithms $\ln^kx$ with $k=1,2,3$, and
(iv) two out of the five polynomials $(1-x)x^k$ with $k=0,\ldots,4$. 
These are $\binom{3}{2}\binom{3}{2}\binom{5}{2}=90$ trial functions with eight
coefficients each, which we pin down using the eight available inputs in
Eqs.~\eqref{guniM4}--\eqref{guniM18}. 
For each coefficient, we determine, from the values thus resulting, the central
value to be half of the sum of the largest and smallest ones and the error to
be half of their difference.
Using all 90 solutions, we find $A_{\mathrm{np}}^{(4)}=-48\times(98.1\pm 5.8)$ and
$B_{\mathrm{np}}^{(4)}=48\times(203.6\pm 32.4)$.
We may considerably improve these results by rejecting 20 unlikely
solutions, involving particularly large coefficients, to obtain 
$A_{\mathrm{np}}^{(4)}=-48\times(97.5\pm 0.6)$ and
$B_{\mathrm{np}}^{(4)}=48\times(207.0\pm 3.0)$.
The former result nicely agrees with the one from
Refs.~\cite{Henn:2019swt,Huber:2019fxe,vonManteuffel:2020vjv},
$A_{\mathrm{np}}^{(4)}=-48(992\pi^6/315+1152\zeta_3^2)=
-48\times 97.75$,
while the latter is new.
We emphasize that our method of computation is completely independent from
Refs.~\cite{Boels:2015yna,Boels:2017skl,Henn:2019swt,Huber:2019fxe,vonManteuffel:2020vjv}.
Our final result for the cusp anomalous dimension through four loops is
\begin{eqnarray}
  \gamma_{\mathrm{cusp}}&=&8\,g^2-26.32\,g^4+190.49\,g^6\nonumber\\
  &&\hspace{-5mm}{}-\Big(1874.86+(97.5\pm0.6)\frac{48}{N_c^2}\Big)g^8
  +{\mathcal O}(g^{10})\,.\quad
\end{eqnarray}

To summarize, using modern computational techniques, we have considerably
advanced our knowledge of the nonplanar sector of $\cN=4$ SYM theory by
studying the universal anomalous dimension of the local, gauge-invariant,
SU(4)-singlet, twist-two operators of definite Lorentz spin $j$ at four loops.
Specifically, we have pushed the state of the art from $j=10$
\cite{Velizhanin:2009gv,Velizhanin:2010ey,Velizhanin:2014zla,Fleury:2019ydf}
to $j=18$ upon providing a first independent confirmation of the recent result
for $j=10$ \cite{Fleury:2019ydf}.
The four new terms for $j=12,\ldots,18$ are all in agreement with the
generic coefficient of $\zeta_5$ already derived in
Ref.~\cite{Velizhanin:2009gv}.
The new information allowed us to uniquely determine also the generic
coefficient of $\zeta_3$, but it does not yet suffice to pin down the generic
expression of the rational term.
However, following Refs.~\cite{Moch:2017uml,Moch:2018wjh}, we managed to find a
rather precise numerical result for the $j\to\infty$ limit of the universal
anomalous dimension by considering the $x\to 1$ limit of the corresponding
splitting function. The result for the cusp anomalous dimension thus obtained
agrees with previous determinations based on very different approaches
\cite{Boels:2015yna,Boels:2017skl,Henn:2019swt,Huber:2019fxe,vonManteuffel:2020vjv}.


Our computations were performed in part with resources provided by the PIK Data Centre in PNPI NRC ``Kurchatov Institute.''
The research of B.A.K. was supported in part by BMBF Grant No.\ 05H18GUCC1 and
DFG Grants No.\ KN~365/13-1 and No.\ KN~365/14-1.
The research of V.N.V. was supported in part by RFBR Grants No.\ 16-02-00943-a, No.\ 16-02-01143-a, and No.\ 19-02-00983-a and a Marie Curie International Incoming Fellowship within the Seventh European Community Framework Programme under Grant No.\ PIIF-GA-2012-331484.

\end{document}